# Optical guiding in meter-scale plasma waveguides

B. Miao*, L. Feder*, J.E. Shrock, A. Goffin, and H.M. Milchberg

*Institute for Research in Electronics and Applied Physics*
*University of Maryland, College Park, MD 20742*

*Abstract*: We demonstrate a new highly tunable technique for generating meter-scale low density plasma waveguides. Such guides can enable electron acceleration to tens of GeV in a single stage. Plasma waveguides are imprinted in hydrogen gas by optical field ionization induced by two time-separated Bessel beam pulses: The first pulse, a $J_0$ beam, generates the core of the waveguide, while the delayed second pulse, here a $J_8$ or $J_{16}$ beam, generates the waveguide cladding. We demonstrate guiding of intense laser pulses over hundreds of Rayleigh lengths with on axis plasma densities as low as $N_{e0} \sim 5 \times 10^{16}$ cm$^{-3}$.

Laser wakefield acceleration (LWFA) of electrons in plasmas has been widely studied in the last several decades[1,2]  While compact, high repetition rate LWFA systems can generate ~10 MeV ultrashort electron bunches[3,4] for numerous applications, high repetition rate generation of multi-GeV bunches is needed for future laser-driven accelerator modules for compact light sources[5] and high energy physics[6,7]. In order to achieve such high energy over relatively short acceleration distances, LWFA requires propagation of high intensity laser pulses over many Rayleigh lengths; some form of optical waveguiding is needed.

Two types of optical guiding in plasmas have been demonstrated: relativistic self-guiding, and guiding in preformed plasma waveguides. In the higher plasma densities required for relativistic self-guiding, electron acceleration is limited to $< \sim 100$ MeV by dephasing and depletion[2,8,9]. Acceleration to multi-GeV energies requires significantly lower plasma densities ($< \sim 5 \times 10^{17}$ cm$^{-3}$) and longer propagation distances for which self-guiding would demand extreme laser power: this limitation motivates use of preformed plasma waveguides.

The first demonstrated plasma waveguide used hydrodynamic cylindrical shock expansion of gas target plasmas driven by inverse bremsstrahlung (IB) plasma heating by a ~100ps axicon-generated Bessel beam pulse; efficient IB heating required a minimum electron density $> \sim 5 \times 10^{18}$ cm$^{-3}$ [10]. Shock expansion established the required concave radial electron density profile for optical guiding: an on-axis electron density minimum $N_{e,0}$ (the 'core') surrounded by the cylindrical shock wall (the 'cladding'). Later work using laser heating of atomic cluster plasmas achieved even lower density waveguides with $N_{e,min} \sim 10^{18}$ cm$^{-3}$[11]. Most recently, the hydrodynamic shock principle was extended to gas targets driven by optical field ionization (OFI) by ultrashort pulses focused by lenses [12–14] or axicons[15].

Another approach adopted in many experiments is the electrical discharge capillary waveguide[16,17]. Here, the guiding channel is determined by the plasma temperature profile established by the heat flow from the hot centre to the cold wall of the capillary. This scheme offered, as originally conceived, the advantage of not requiring a laser.  Recently, however, an experiment demonstrating 8 GeV electron acceleration in a 20 cm discharge capillary[18] required auxiliary laser heating to reduce the waveguide core density. In addition, their high localized heat load makes discharges an unlikely component of future high repetition rate accelerators. As the technology, cost, and efficiency improves for high repetition rate, high energy ultrashort pulse lasers[19], plasma waveguides in future accelerators will likely be *laser*-driven.

---

* contributed equally to this work

Laser-driven OFI waveguides relying on hydrodynamic shock expansion are limited to low electron temperatures comparable to the electron ponderomotive energy in the laser field at the ionization threshold of the gas[20]. At the hydrogen ionization threshold of $\sim 10^{14}$ W/cm$^2$, $T_e < \sim 10$ eV for a $\lambda = 1$ µm laser, resulting in weak expansion leading to low shock walls with limited optical confinement. OFI heating can be enhanced by using a gas with higher ionization potential such as helium[21], but for a fixed pulse width, the laser energy must be increased by $> 100 \times$ compared to hydrogen.

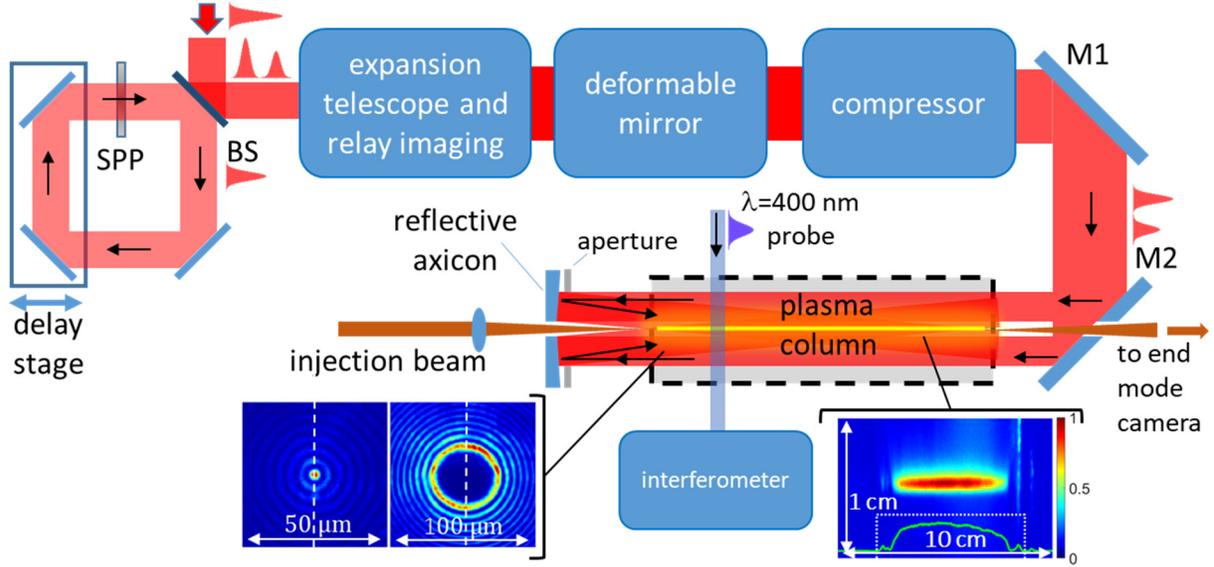

**Figure 1.** Experimental setup. The ~200 ps uncompressed pulse from a Ti:Sapphire amplifier is split by a non-polarizing beamsplitter (BS). The transmitted pulse is converted to a $(0, q)$ beam by a segmented $q^{th}$ order spiral phase plate (SPP). After beam expansion and relay imaging, phase front correction using a deformable mirror, and re-compression, the beam is focused by a reflective axicon in either backfill hydrogen gas or over a supersonic gas jet. The counter-propagating guided beam is injected into the waveguide through a hole in the axicon and imaged after guiding. Also shown: frequency-doubled (λ=400 nm, 70 fs) interferometric probe pulse. *Left bottom inset*: images of $J_0$ and $J_{16}$ beams in focus. *Right bottom inset*: $J_0$-pulse-induced plasma fluorescence 3.5 mm above the 5 cm supersonic H$_2$ jet.

In this paper, we present a new approach to OFI-generated waveguides and demonstrate high intensity optical guiding up to hundreds of Rayleigh ranges. Our approach decouples the generation of the needed electron density profile from plasma heating and does not exclusively rely on hydrodynamic expansion. It leverages the idea that OFI can effectively imprint prescribed electron density profiles while minimally heating them.

As detailed in Fig. 1, an amplified and stretched (to ~200ps) linearly polarized λ=800nm pulse from a Ti:Sapphire laser is split at an 80/20 non-polarizing beamsplitter, with 80% of the energy entering a delay ring containing a transmissive spiral phase plate (SPP) to impose either a $16\pi$ ($q = 8$) or $32\pi$ ($q = 16$) azimuthal phase screw on the beam. The recombined output beam is a $(p = 0, m = 0)$ Laguerre super-Gaussian beam pulse followed, at a delay $\tau_d = 1 - 3$ ns, by a $(p = 0, m = q)$ pulse, where $p$ and $m$ are radial and azimuthal mode indices. The beam is then expanded and relay imaged, phase front-corrected, recompressed to 50-100 fs FWHM, and directed to a 50.8mm diameter reflective axicon in an experimental chamber. The $4 \times$ relay imaging conveys the plane of the spiral phase plate to the axicon. Phase front correction improves the quality of the $q = 8$ or $q = 16$ beams, employing a deformable mirror using a phase front retrieval technique developed by our group[22]. The reflective axicon forms a double-pulsed Bessel-Gauss beam, identified here using only the Bessel notation as a $J_0$ pulse followed $\tau_d = 1 - 3$ ns later by a $J_q$ pulse. The $J_0$ pulse forms the core of the plasma waveguide by OFI; the $J_q$ pulse forms the cladding. Here we use an H2 backfill target or an elongated H2 gas jet.

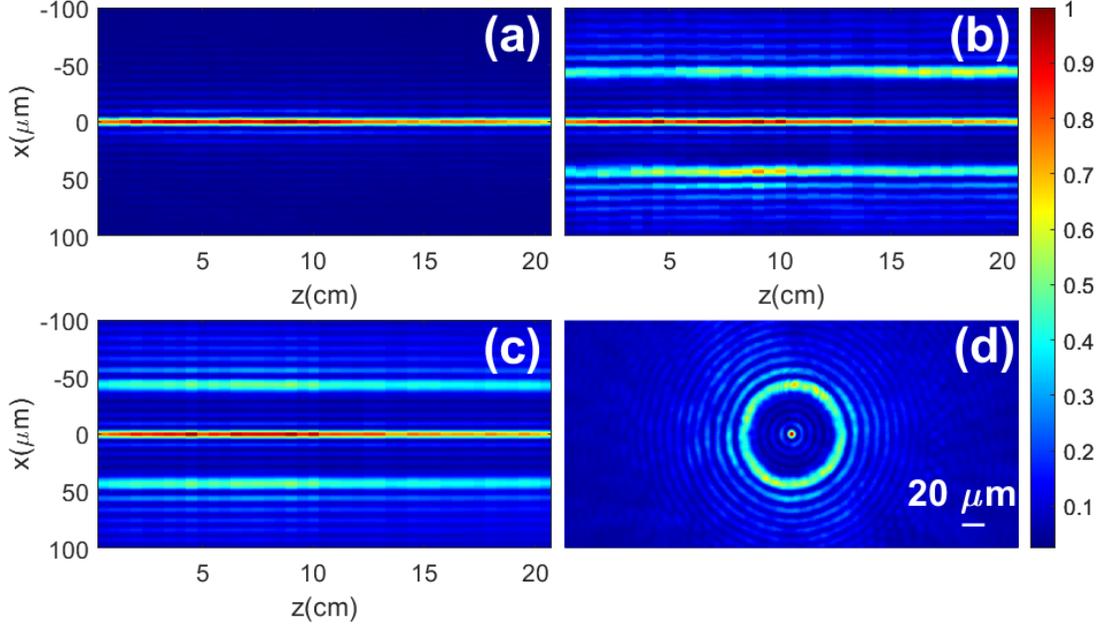

**Figure 2. (a)** Optical axis scan of $J_0$ focal spot profiles with planar cut shown in Fig. 1 inset. **(b)** Optical axis scan of combined $J_0$ and $J_{16}$ beam, with planar cut shown in Fig. 1 inset. **(c)** Azimuthal average of (b). **(d)** Combined $J_0 + J_{16}$ beam profile. $\gamma = 3\,°$

In our experiments, the Bessel beam depth of focus, or line focus length, is $L \approx (R - a)/\tan \gamma \gtrsim 15$ or 30 cm, where $R = 2$ cm is the radius of the outer aperture near the axicon surface, $a$ is the radius of the axicon central hole, $\gamma = 2\alpha$ is the Bessel beam ray approach angle to the optical axis, and $\alpha = 1.5°$ or $3°$ is the axicon base angle. The inner and outer edges at $a$ and $R$ set up sharp transitions of Bessel beam intensity at the extremes of the focal line, preventing excessive axial taper of the plasma there. In vacuum, the central peak radius of the $J_0$ beam was $r_0 = 2.405/k\sin\gamma = 6$ or 3 μm, where $k$ is the laser vacuum wavenumber. The initial $J_0$ pulse fully ionizes the hydrogen gas target, creating a long thin plasma column which expands cylindrically outward, leaving a region of low electron density near the optical axis, creating the "core" of the plasma waveguide. The delayed $J_q$ pulse generates a long cylindrical plasma shell[23–25] around the expanded column: this serves as the lower refractive index "cladding". The intense pulse to be guided is focused through a hole in the axicon and coupled into the waveguide.

An important first requirement is for the $J_0$ pulse to fully ionize the hydrogen gas target to prevent ionization defocusing of the guided pulse. We found that an optical axis-averaged peak intensity of $\sim 4 \times 10^{14}$ W/cm$^2$ in the $J_0$ central maximum was sufficient to fully ionize the hydrogen gas and clamp the electron temperature along the focal line sections used in our experiments, despite the axial variation in the intensity dictated by the mapping of the beam profile to the focal line $I(z) = I_0 z \exp(-(z/w)^n)$ ($z = r/\tan\alpha, r \geq a, n \approx 6$). Figures 2(a) and 2(b) plot cuts of $J_0$ alone and combined $J_0 + J_{16}$ along the white-dashed plane edges in Fig. 1. These composite images were obtained by scanning a 10× microscope objective along $z$. The profiles remain relatively uniform over tens of centimeters, with $z$ variations mostly from radial variations of the input beam. Figure 2(c) shows 2(b) azimuthally averaged, and Fig. 2(d) shows a transverse cut across $z$ of the combined beams.

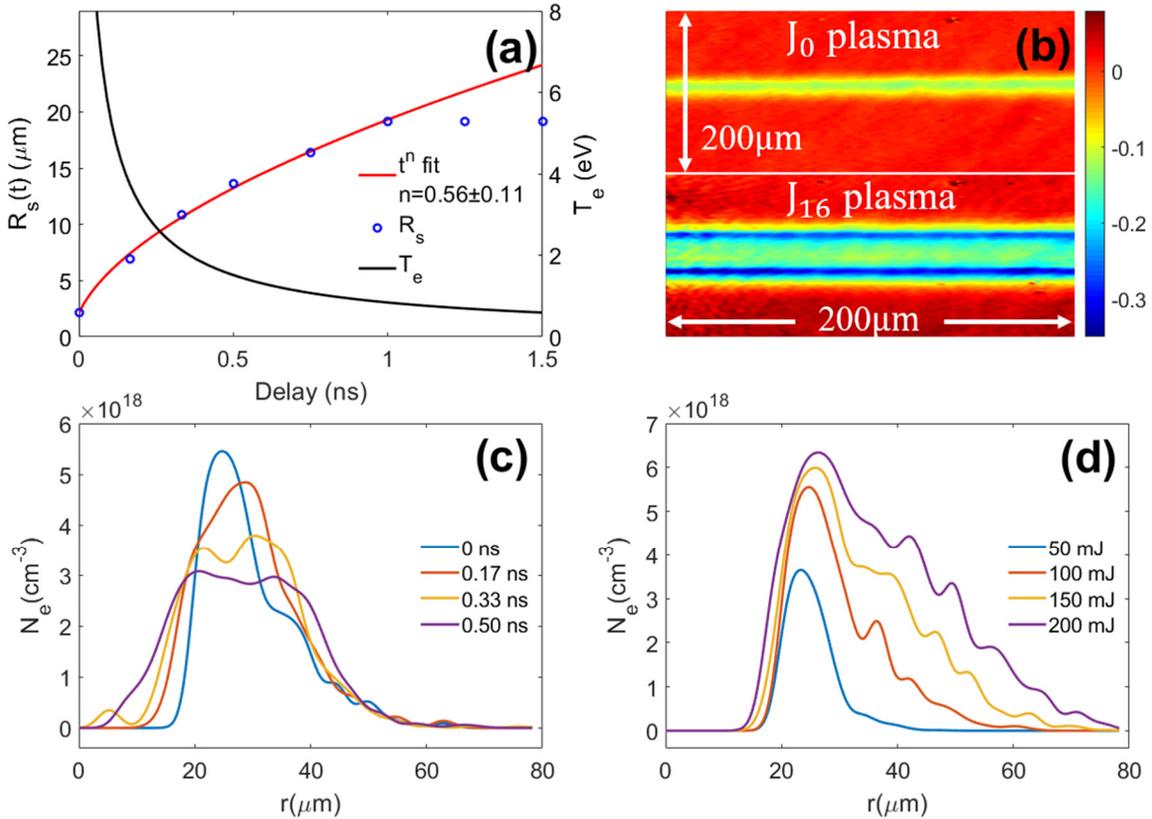

**Figure 3. (a)** Radius of blast wave expansion of $J_0$ plasma (best fit $n = 0.56 \pm 0.11$) and extracted electron temperature vs. time. Hydrogen backfill 50 torr. **(b)** Interferometric phase shift images of $J_0$ and $J_{16}$ plasmas 10ps after generation. Hydrogen backfill 100 torr. **(c)** Time evolution of $J_{16}$-generated annular electron density profile. The plasma pressure gradient mainly drives filling of the central hole. Pulse energy=100 mJ. **(d)** Pulse energy dependence of $J_{16}$-generated annular electron density profiles, with saturation for >50 mJ. The modulations at larger radius are from ionization by outer rings of the $J_{16}$ pulse.

After OFI by the $J_0$ pulse, the electrons thermalize over the electron-electron collision time $\tau_{ee} <$ 1 ps, for $N_e \sim 10^{18}$ cm$^{-3}$ and initial electron temperature $T_{e0} < \sim 10$eV, leaving a warm plasma column surrounded by cold, neutral hydrogen gas. The plasma column rapidly cools as it expands outward as a cylindrical blast expansion $R_s(t) = \xi_0 (\varepsilon_0/\rho)^{1/4} t^{1/2}$, where $R_s$ is the radial position of the outer plasma boundary, $\varepsilon_0$ ($\propto T_{e0}$) is the laser energy deposition per unit length, $\rho$ the initial mass density of the hydrogen gas, and $\xi_0$ is a dimensionless prefactor of order unity [26]. Typical plots of $R_s(t)$ and associated density profiles are shown in Fig. 3(a) and Fig. 4(a) for peak $J_0$ intensity $10^{15}$ W/cm$^2$, and backfill hydrogen pressure of 50 Torr. The electron temperature $T_e$ is determined from the ion acoustic speed as $dR_s/dt \approx c_s = (\gamma_c T_e/m_i)^{1/2}$, where $\gamma_c$ is the specific heat ratio and $m_i$ is the proton mass. It is seen that after a few nanoseconds, the expansion stagnates as the plasma cools to < 1 eV, with the central plasma density reduced by ~10 × and the column expanded by ~3 − 5 ×. Two-color interferometry and integration of the total phase shift from the plasma column indicate that there is no recombination on this time scale. The weak $(T_{e0})^{1/4}$ dependence of the plasma expansion ensures that moderate axial variations in Bessel beam intensity negligibly affects the local axial uniformity of the waveguide plasma, as seen in Fig 3(b).

After the $J_0$ pulse has formed the core of the plasma waveguide, the delayed $J_q$ pulse forms the cladding. The index $q$ of the high order beam and its delay are chosen so that it propagates through the expanded $J_0$ plasma column, forming a high intensity annulus at its edge, ionizing the neutrals there. The annular plasma formed by a $J_{16}$ pulse alone is shown in Fig. 3(c), as function of delay after generation at $t = 0$, and in Fig. 3(d) at $t = 0$ as a function of $J_{16}$ pulse energy. It is seen that the annular plasma cladding decays over ~0.5 ns, mainly by filling in the central hole, and that the cladding height saturates at >50 mJ.

The effect of the $J_0$ –generated plasma on the propagation of the $J_q$ pulse was calculated using a simulation of ultrashort pulse Bessel beam propagation including OFI [27]. Figure 4(b) shows the $J_{16}$ pulse-generated electron density profiles with and without the $J_0$-generated core plasma present. The effect of the preformed core plasma on the $J_{16}$-pulse propagation and induced OFI profile is seen to be negligible. In general, unless the core plasma density approaches the effective critical density $N_{cr,eff} = N_{cr}\sin^2\gamma \sim 0.5 - 2 \times 10^{19}$ cm$^{-3}$ experienced by oblique Bessel beam rays, it has a negligible effect on the $J_q$ –induced ionization profile.

The two Bessel-pulse method of plasma waveguide generation enables very wide tuning of the plasma waveguide transverse profile and its guided modes through control of the hydrogen density, the order $q$ of the second pulse, the Bessel beam axis approach angle $\gamma$ ($= 2\alpha$), the energy and pulsewidth of the two pulses, and the time delay between them. The plots in Fig. 4(c)-(f) illustrate this flexibility. Here, the delay between the $J_0$ and $J_q$ pulses is $\tau_d = 2.4$ ns. In general, the desired guided optical mode dictates the required waveguide profile, with leaky mode or beam propagation method simulations[27] providing the best design guidance. A good design approximation, however, recognizes the step-like profiles of our two-pulse OFI waveguides: a relatively flat core and sharply rising cladding, for which the fundamental channel mode radius is given by[28] $w_{ch} \approx a(0.6484 + 1.619V^{-3/2} + 2.879V^{-6} + \cdots)$, where $a$ is the core radius and the step index fibre parameter is $V = ka(n_{core}^2 - n_{cladding}^2)^{1/2} = a(4\pi r_e \Delta N_e)^{1/2}$, where $\Delta N_e = N_e^{cladding} - N_e^{core}$, $r_e$ is the classical electron radius, and $k$ is the laser vacuum wavenumber. Larger guided modes can be supported by smaller core-cladding differences $\Delta N_e$, which require lower hydrogen density $N_0$ (since under OFI, $\Delta N_e$ is roughly proportional to $N_0$).

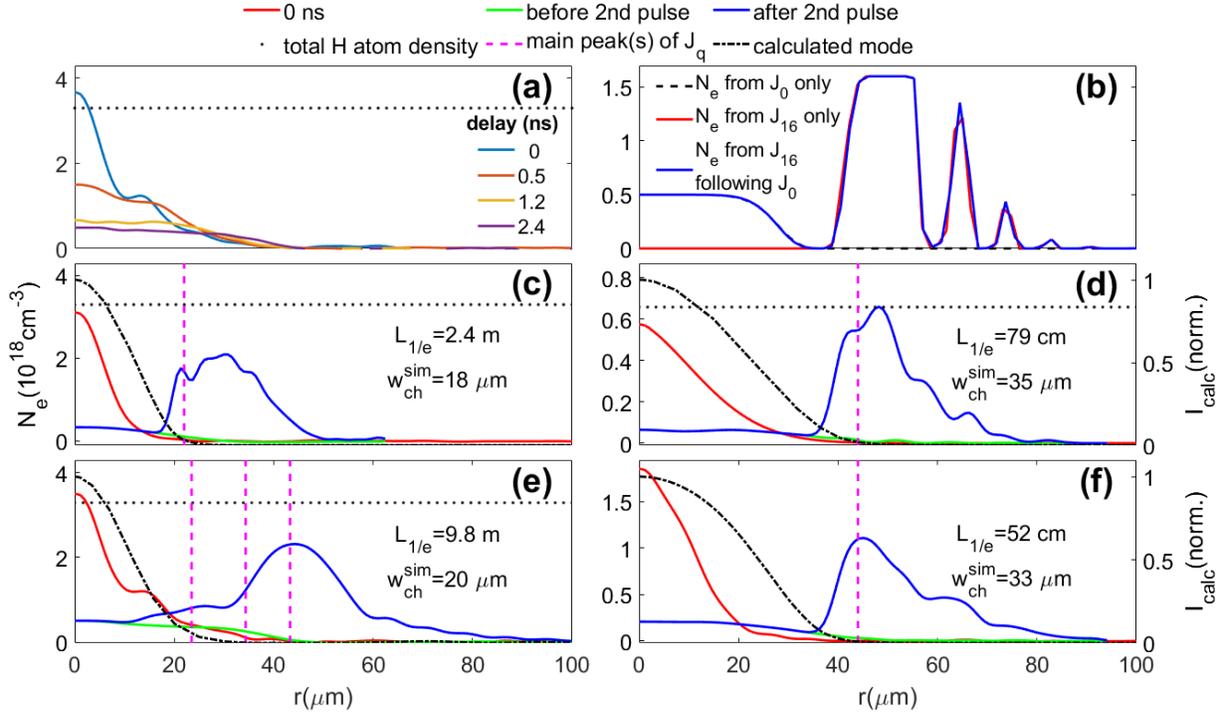

**Figure 4.** Transverse electron density profiles in backfill and gas jet hydrogen plasmas. Where shown, the magenta dashed lines indicate maxima of $J_q$ rings. **(a)** $J_0$-generated 'core' electron density profiles vs. time delay **(b)** $J_{16}$ pulse propagation simulations of plasma profile generation with and without $J_0$-generated core plasma present. $\alpha = 1.5°$, $(c) - (f)$: Electron density profiles measured ~10 ps after the $J_0$ pulse (red curves), and 10 ps before and after the $J_q$ pulse (green curves and blue curves). **(c)** 100 mJ ($J_0$) +235 mJ ($J_{16}$), $\alpha = 3°$, $H_2$ backfill pressure P=50 torr. Waveguide length $L = 15$ cm; **(d)** 78 mJ ($J_0$)+250 mJ ($J_{16}$), $\alpha = 1.5°$, $H_2$ backfill pressure P=10 torr. Waveguide length $L = 30$ cm; **(e)** 100 mJ ($J_0$)+235 mJ ($J_8$), $\alpha = 1.5°$, $H_2$ backfill pressure P=50 torr. Waveguide length $L = 30$ cm; **(f)** $H_2$ gas jet, $P \approx 30$ torr: 87 mJ ($J_0$)+235 mJ ($J_{16}$), $\alpha = 1.5°$, $q=16$. Waveguide length $L = 5$ cm.

This mode scaling is illustrated in Fig. 4, where the $H_2$ density in Fig. 4(c) is chosen to be $5 \times$ higher than in Fig. 4(d). Both experiments use a $J_{16}$ pulse to generate the waveguide cladding, the approach angle decreasing from $\gamma = 6°$ in 3(c) to $\gamma = 3°$ in 3(d). This has the effect of widening the initial $J_0$-generated fully ionized core plasma and moving the cladding peak from a radial position of ~30 μm out to ~50 μm. In both cases, the core plasma remains unaffected by the second pulse, as seen by the merging of the blue and green curves toward the waveguide axis. In Fig. 4(c), $\Delta N_e \sim 1.3 \times 10^{18}$ cm$^{-3}$, $a$~20μm, and $V$~4.3, predicting $w_{ch}$~17μm, while in Fig. 4(d), $\Delta N_e$~0.5 × $10^{18}$ cm$^{-3}$, $a$~40μm, and $V$~5.4 give $w_{ch}$~32 μm. We also plot the fundamental modes and their $1/e$ intensity attenuation lengths $L_{1/e}$ computed using our leaky mode solver[27,29], showing good agreement with the step index values. As an example of the control available for sculpting transverse profiles, Fig. 4(e) shows use of a $J_8$ pulse at approach angle $\gamma = 3°$. Here, the peaks of the first and second $J_8$ rings overlap with ~50% ionized hydrogen in the expanded plasma from the $J_0$ pulse, generating an inner cladding bump, while the third ring overlaps neutral hydrogen gas on the plasma periphery[27], generating a higher outer cladding bump. Considering this as a step index profile to the first cladding bump (with $\Delta N_e$~0.5 × $10^{18}$ cm$^{-3}$, $a$~15μm, and $V$~2.0), gives $w_{ch}$~19 μm, in good agreement with the leaky mode solver.

We have presented results thus far for waveguides generated in $H_2$ backfill gas. For LWFA, however, distortion-free injection and guiding of high intensity pulses demands that the waveguide entrance and exit be free of neutral gas. For this purpose, we have developed a supersonic gas jet producing uniform low pressure flows. The inset to Fig. 1 shows uniform fluorescence from $J_0$ pulse-induced OFI of a 5 cm long $H_2$ gas sheet, 3.5 mm above the nozzle orifice where the pressure is ~30 Torr. The 15 cm long Bessel beam focus overfills the gas sheet. For jets longer than ~ 5 cm, the gas nozzle interferes with $J_q$ formation in the focus; we have recently developed an optical fix for this limitation[30]. Figure 4(f) shows results from 2-pulse OFI waveguide generation in the gas jet, where a step-like guide is clearly formed. Applying the step index analysis (using $\Delta N_e \sim 0.75 \times 10^{18}$ cm$^{-3}$, $a \sim 40$μm, and $V \sim 2.0$) gives $w_{ch} \sim 30$ μm, in reasonable agreement with the leaky mode computation.

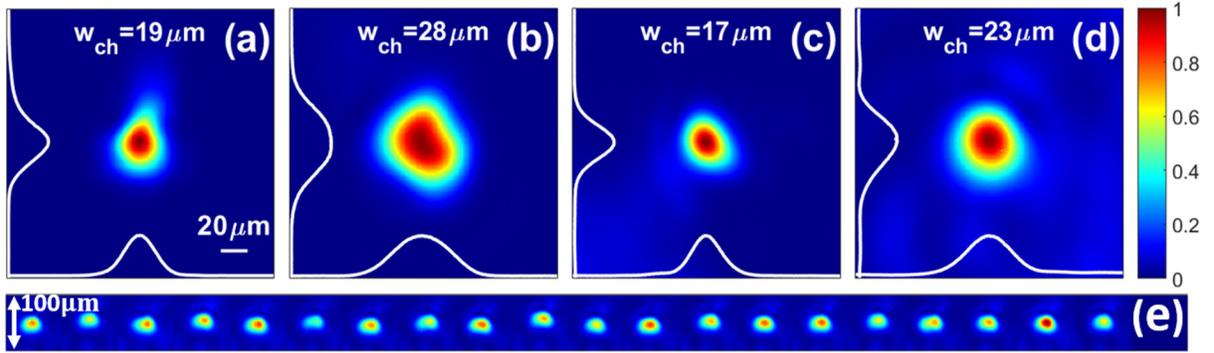

**Figure 5. (a)(b)(c)** Low intensity exit modes from $L$ =15, 30, and 30 cm long plasma waveguides (number of Rayleigh ranges $L/z_0 \sim$ 110, 100, and 260 (for λ=800nm)) under conditions of Fig. 4(c)(d)(e), for injection of ~100 μJ, $\lambda = 400$ nm pulse. **(d)** Gas jet experiment: high intensity exit mode of a 5 cm long waveguide generated in $H_2$ gas jet. Injected pulse 40 mJ, λ=800 nm, 50 fs FWHM, with ~50% coupling efficiency, giving guided intensity ~$6 \times 10^{16}$ W/cm² over $L/z_0 \sim 25$ Rayleigh ranges **(e)** Guided modes from 20 successive gas jet shots for conditions of (d). Variations are mainly from injected beam pointing fluctuations.

Guided modes for the $L = 5$ cm $- 30$ cm waveguides of Fig. 4 are shown in Fig. 5, where the modes of Fig. 5(a)-(d) correspond to the conditions of Fig. 4(c)-(f) (and (5(d)(e) and 4(f) are for the jet experiment). For these spot sizes and guide lengths, the number of Rayleigh lengths of guiding (at λ=800nm) ranges over $L/z_0$ ~25 to ~260. In all cases, backfill and gas jet, the guided pulse was injected into the waveguide within ~10 ps after the cladding-generating $J_q$ pulse. Even though the core expands over ~ 2 ns and the cladding collapses in ~0.5 ns, as seen in Figs. 2 and 3, this does not limit the length of fixed-parameter waveguides produced using our method: any guided pulse can propagate a couple of nanoseconds behind the $J_0$ pulse and immediately behind the traveling wave cladding formation by the $J_q$ pulse. As can be seen in Fig. 5, the guided modes are slightly asymmetric owing to imperfectly corrected azimuthal intensity variations in the $J_q$ beam. This leads to slight azimuthal guide variations, as we have determined using a new quasi-2D Abel inversion algorithm employing the measured $J_q$ beam asymmetry[27]. The guided mode radii $w_{ch}$ in the figure are from 2D Gaussian fits to the measured mode, and agree reasonably well with the computed modes, where the electron density profiles were azimuthally averaged from the extracted 2D profiles.

The guided mode in the hydrogen jet plasma waveguide (Fig. 5(d)) has peak intensity ~$6 \times 10^{16}$ W/cm², based on ~50% coupling efficiency, where the injected laser energy was limited to 40 mJ by our system's available laser energy. There was no evidence of additional ionization by the guided pulse, consistent with our density profile measurements showing fully ionized waveguide cores. Exit modes from a sequence of 20 successive gas jet shots is shown in Fig. 5(e). Pointing fluctuations of the injection pulse are mainly responsible for the variation shown.

Based on the results presented here, we can estimate potential electron acceleration driven by a more powerful laser than ours, which would enable longer waveguides and higher injection energies. As an example, the waveguide in Fig. 4(d) has a core electron density of $N_{e0} \sim 5 \times 10^{16} cm^{-3}$ and $w_{ch} = 35\ \mu m$. This corresponds to a LWFA dephasing length $L_{deph} \approx 1.3$ m [2], which exceeds $L_{1/e} \sim 0.8$ m for this guide. A resonant pulse with $a_0=1$ (for which $L_{deph} \sim L_{depletion}$) and $\tau \approx (\lambda_p/c)/(\pi\sqrt{2}) \approx 110$ fs guided over $L_{1/e}$ would give an energy gain $\Delta E_{accel} \sim 13$ GeV [9] for laser energy 4.5 J and peak power $\sim 40$ TW. As seen from the waveguide examples in Fig. 4, wide control over $w_{ch}$ and $L_{1/e}$ is possible. Increasing the guided spot size to $w_{ch} \approx \lambda_p/2 \sim 75\ \mu m$ for the same $N_{e0}$, and extending the corresponding waveguide so that $L_{1/e} > L_{deph} = 2.1$ m, would yield $\Delta E_{accel} \sim 34$ GeV for laser energy 21 J and peak power 190 TW, with $\sim 5 \times$ more charge accelerated. For this case, the laser energy cost for the waveguide would be $< \sim 5$ J.

In conclusion, we have presented a new, highly tunable technique for generating long, low loss plasma waveguides for a wide range of plasma densities and mode sizes. We have demonstrated guiding in waveguides up to 30 cm and over 250 Rayleigh lengths, with a guided intensity limited only by our available laser energy. This technique is ideal for multi-GeV laser wakefield acceleration.

*Acknowledgements.* The authors thank M. Tomlinson for technical assistance, and A. Picksley, A. Ross, and S. Hooker for useful discussions. This research is supported by the US Department of Energy (DESC0015516) and the National Science Foundation (PHY1619582).

**References**

[1] T. Tajima and J. M. Dawson, "Laser Electron Accelerator," *Phys. Rev. Lett.* **43**, 267 (1979).

[2] E. Esarey, C. B. Schroeder, and W. P. Leemans, "Physics of laser-driven plasma-based electron accelerators," *Rev. Mod. Phys.* **81**, 1229 (2009).

[3] F. Salehi, A. J. Goers, G. A. Hine, L. Feder, D. Kuk, B. Miao, D. Woodbury, K. Y. Kim, and H. M. Milchberg, "MeV electron acceleration at 1 kHz with <10 mJ laser pulses," *Opt. Lett.* **42** (2017).

[4] D. Guénot, D. Gustas, A. Vernier, B. Beaurepaire, F. Böhle, M. Bocoum, M. Lozano, A. Jullien, R. Lopez-Martens, et al., "Relativistic electron beams driven by kHz single-cycle light pulses," *Nat. Photonics* **11**, 293 (2017).

[5] F. Albert and A. G. R. Thomas, "Applications of laser wakefield accelerator-based light sources," *Plasma Phys. Control. Fusion* **58**, 103001 (2016).

[6] R. W. U. Dorda, R. Assmann, M. Ferrario, E. Gschwendtner, B. Holzer, A. Mosnier, J. Osterhoff, A. Specka, A. Walker, Ed., "3rd European Advanced Accelerator Concepts workshop (EAAC2017)," in *3rd Eur. Adv. Accel. Concepts Work.*, (La Biodola, Isola

d'Elba, 2018).

[7] Evgenya I. Simakov ; Nikolai Yampolsky ; Kent P. Wootton, Ed., "2018 IEEE Advanced Accelerator Concepts Workshop (AAC)," in *2018 IEEE Adv. Accel. Concepts Work.*, (IEEE, Breckenridge, Colorado, USA, 2018).

[8] W. Lu, C. Huang, M. Zhou, W. B. Mori, and T. Katsouleas, "Nonlinear theory for relativistic plasma wakefields in the blowout regime.," *Phys. Rev. Lett.* **96**, 165002 (2006).

[9] W. Lu, M. Tzoufras, C. Joshi, F. S. Tsung, W. B. Mori, J. Vieira, R. A. Fonseca, and L. O. Silva, "Generating multi-GeV electron bunches using single stage laser wakefield acceleration in a 3D nonlinear regime," *Phys. Rev. ST Accel. Beams* **10**, 61301 (2007).

[10] C. G. Durfee and H. M. Milchberg, "Light pipe for high intensity laser pulses," *Phys. Rev. Lett.* **71**, 2409 (1993).

[11] V. Kumarappan, K. Y. Kim, and H. M. Milchberg, "Guiding of Intense Laser Pulses in Plasma Waveguides Produced from Efficient, Femtosecond End-Pumped Heating of Clustered Gases," *Phys. Rev. Lett.* **94**, 205004 (2005).

[12] N. Lemos, T. Grismayer, L. Cardoso, G. Figueira, R. Issac, D. A. Jaroszynski, and J. M. Dias, "Plasma expansion into a waveguide created by a linearly polarized femtosecond laser pulse," *Phys. Plasmas* **20**, 063102 (2013).

[13] R. J. Shalloo, C. Arran, L. Corner, J. Holloway, J. Jonnerby, R. Walczak, H. M. Milchberg, and S. M. Hooker, "Hydrodynamic optical-field-ionized plasma channels," *Phys. Rev. E* **97**, 053203 (2018).

[14] N. Lemos, L. Cardoso, J. Geada, G. Figueira, F. Albert, and J. M. Dias, "Guiding of laser pulses in plasma waveguides created by linearly-polarized femtosecond laser pulses," *Sci. Rep.* **8**, 3165 (2018).

[15] R. J. Shalloo, C. Arran, A. Picksley, A. von Boetticher, L. Corner, J. Holloway, G. Hine, J. Jonnerby, H. M. Milchberg, et al., "Low-density hydrodynamic optical-field-ionized plasma channels generated with an axicon lens," *Phys. Rev. Accel. Beams* **22**, 041302 (2019).

[16] Y. Ehrlich, C. Cohen, A. Zigler, J. Krall, P. Sprangle, and E. Esarey, "Guiding of High Intensity Laser Pulses in Straight and Curved Plasma Channel Experiments," *Phys. Rev. Lett.* **77**, 4186 (1996).

[17] A. Butler, D. J. Spence, and S. M. Hooker, "Guiding of High-Intensity Laser Pulses with a Hydrogen-Filled Capillary Discharge Waveguide," *Phys. Rev. Lett.* **89**, 185003 (2002).

[18] A. J. Gonsalves, K. Nakamura, J. Daniels, C. Benedetti, C. Pieronek, T. C. H. de Raadt, S. Steinke, J. H. Bin, S. S. Bulanov, et al., "Petawatt Laser Guiding and Electron Beam Acceleration to 8 GeV in a Laser-Heated Capillary Discharge Waveguide," *Phys. Rev. Lett.* **122**, 084801 (2019).

[19] "Advanced Solid State Lasers," in *Laser Congr. 2019 (ASSL, LAC, LS&C)*, (OSA, Vienna, Austria, 2019).

[20] P. B. Corkum, N. H. Burnett, and F. Brunel, "Above-Threshold Ionization in the Long-Wavelength Limit," *Phys. Rev. Lett.* **62**, 1259 (1989).

[21] I. Pagano, J. Brooks, A. Bernstein, R. Zgadzaj, J. Leddy, J. Cary, and M. C. Downer, "Low Density Plasma Waveguides Driven by Ultrashort (30 fs) and Long (300 ps) Pulses for Laser Wakefield Acceleration," in *2018 IEEE Adv. Accel. Concepts Work.*, (2018).

[22] Bo Miao *et al*. (to be submitted)

[23] J. Fan, E. Parra, I. Alexeev, K. Y. Kim, H. M. Milchberg, L. Y. Margolin, and L. N. Pyatnitskii, "Tubular plasma generation with a high-power hollow Bessel beam," *Phys. Rev. E* **62**, R7603–R7606 (2000).

[24] W. D. Kimura, H. M. Milchberg, P. Muggli, X. Li, and W. B. Mori, "Hollow plasma channel for positron plasma wakefield acceleration," *Phys. Rev. Spec. Top. - Accel. Beams* **14**, 041301 (2011).

[25] S. Gessner, E. Adli, J. M. Allen, W. An, C. I. Clarke, C. E. Clayton, S. Corde, J. P. Delahaye, J. Frederico, et al., "Demonstration of a positron beam-driven hollow channel plasma wakefield accelerator," *Nat. Commun.* **7**, 11785 (2016).

[26] T. R. Clark and H. M. Milchberg, "Time- and Space-Resolved Density Evolution of the Plasma Waveguide," *Phys. Rev. Lett.* **78**, 2373 (1997).

[27] "Supplemental material for 'Optical guiding in meter-scale plasma waveguides.'"

[28] D. Marcuse, "Loss Analysis of Single-Mode Fiber Splices," *Bell Syst. Tech. J.* **56**, 703–718 (John Wiley & Sons, Ltd, 1977).

[29] T. R. Clark and H. M. Milchberg, "Optical mode structure of the plasma waveguide," *Phys. Rev. E* **61**, 1954 (2000).

[30] J. E. Shrock *et al*. (to be submitted)

# Supplemental material for "Optical guiding in meter-scale plasma waveguides"

## 1. Interferometric measurement of plasma density profiles

***1D extraction for azimuthally symmetric profiles.*** Transverse profiles of electron density were extracted from interferograms generated by an ultrashort $\lambda = 400$ nm, 70 fs probe pulse, imaged from the plasma through a femtosecond shearing interferometer and onto a CCD camera. Each profile was constructed using 200 shots (frames). The analysis steps are as follows. First, the 2D interferometric phase shift $\Delta\Phi(y, z)$ was extracted from each frame using standard techniques[1], where $z$ and $y$ are coordinates along and transverse to the optical axis. In each column (fixed $z$), background subtraction was performed by fitting the phase in the plasma-free part of the frame to a polynomial and interpolating over the full frame. An average phase shift $\overline{\Delta\Phi(y)}$ was then computed over 500 columns of the extracted phase and then over 200 shots, where the plasma position and phase shift profile is very stable from shot to shot. This procedure resulted in phase shift resolution of ~ 0.4 mrad, enabling extraction of electron densities as low as $4 \times 10^{16} \text{cm}^{-3}$ (assuming a minimum signal-to-noise ratio as 20 dB). $\overline{\Delta\Phi(y)}$ was then Abel inverted to recover the plasma density.

***Quasi-2D extraction for azimuthally asymmetric profiles.*** For the symmetric plasmas generated by $J_0$-pulses, the above procedure was sufficient. However, for plasmas generated by a $J_q$ pulse, $\Delta\Phi(y, z)$ was often asymmetric. We attributed this apparent azimuthal asymmetry to asymmetry in the intensity profile of the $J_q$ beam.

While the standard Abel inversion assumes azimuthal symmetry of the density profile to be measured, it can be used in modified form for a profile that is a separable function of radial and azimuthal coordinates ($r$ and $\phi$). Here, we use a separate measurement of the $J_q$ intensity profile to inform the extraction algorithm. First, since the intensity of the $J_q$ beam is near zero inside its first ring, we assume that there is no further ionization there and subtracted the phase shift $(\Delta\Phi(y, z))_{J_0}$ of the $J_0$ plasma from the total phase shift from the two pulses $(\Delta\Phi(y, z))_{J_0+J_q}$. We then assume that the plasma profile follows the same asymmetry as the $J_q$ intensity profile and that its radial dependence is smooth and non-negative. A modified Abel Inversion was performed with these assumptions, and then the pre-existing $J_0$ plasma was added back into the electron density profile.

To account for the azimuthal asymmetry in the density profile, we added azimuthal terms to the Abel Inversion with coefficients derived from the $J_q$ profile, and assume that the phase profile to be extracted can be expressed as $F(r, \phi) = f(r) \sum_{m=0}^{n} c_m \cos(m\phi + \Delta\phi_m)$. Then the plasma density profile is extracted by $\text{argmin}_F \int |\mathcal{A}[F] - \Delta\Phi(y)|^2 dy$, subjected to $F(r, \phi) \geq 0$ and $f(r < r_{min}) = 0$ ($r_{min} = 10 \mu m$). Here $\mathcal{A}[F]$ is the Abel projection operator and $\Delta\Phi(y)$ is measured phase shift in the experiment. The radial profile $f(r)$ was fit by a Gaussian basis set to improve smoothness and account for the limited resolution in interferometry, as detailed in [2]. The $\Delta\phi_m$ terms are determined by fitting the measured $J_q$ focus profile azimuthally, as shown in Fig. S1. In the optimization we keep azimuthal terms up to $m = 6$. The optimization used the built-in MATLAB function *fmincon*, and $f(r)$ was initialized with random noise. The criterion of convergence was set as the normalized error $\epsilon = \int |\mathcal{A}[F] - \Delta\Phi(y)|^2 dy / \int |\Delta\Phi(y)|^2 dy < 0.001$. To further verify uniqueness of the solution, we synthesized $F(r,\phi)$ with randomly shaped ring profiles and azimuthal terms up to $m = 6$, and constructed Abel transforms of these synthesized profiles as measured phase shift profiles. Reconstruction of these dummy phase shift profiles using this method showed correct results upon convergence when $f(r)$ was initialized by random noise. Therefore, we conclude that uniqueness was achieved.

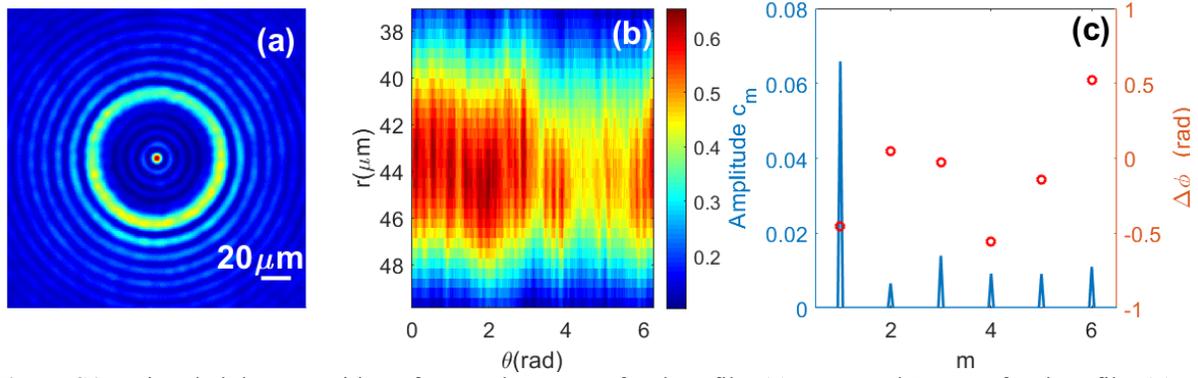

**Figure S1**. Aziumthal decomposition of a sample $J_0 + J_{16}$ focal profile. **(a)** Measured $J_0 + J_{16}$ focal profile. **(b)** Radial lineouts of main $J_{16}$ ring vs. angle. The colorbar shows intensity normalized to the central $J_0$ maximum of (a). **(c)** Azimuthal decomposition of the radially averaged ring intensity of (b).

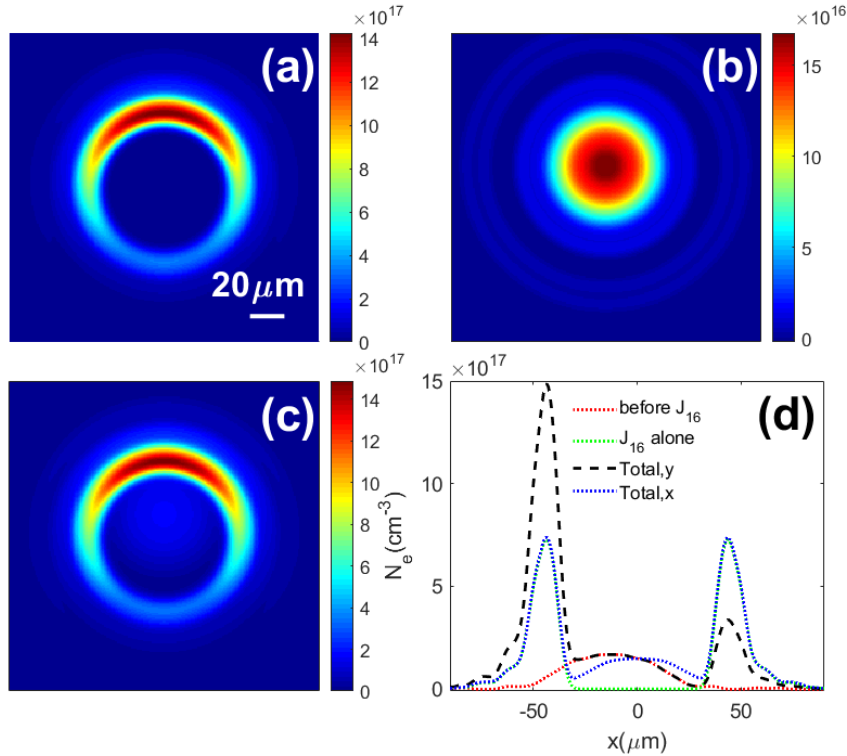

**Figure S2.** Full 2D extraction of the electron density profile shown in Fig. 4(f) of the main paper. **(a)** Electron density profile from the $J_{16}$ pulse only. **(b)** Electron density from the $J_0$ pulse, right before the arrival of the $J_{16}$ pulse. **(c)** Total electron density after the $J_{16}$ pulse, **(d)** Lineouts of panels (a)-(c).

While this method produces 2D plasma density profiles, azimuthally averaged lineouts were plotted in the main paper. The focal profile was measured along the full focal line, and the average $\Delta\phi_m$ was used in the minimization. An example of the full 2D electron density profile is shown in Fig. S2. The asymmetry in Fig. S2(a) and S2(c) is attributed to non-uniformity of the $J_{16}$ beam.

## 2. Calculation of quasi-bound modes

As discussed in detail in [3], plasma waveguides with a finite thickness cladding have modes that are quasi-bound or leaky. A sufficiently well-bound leaky mode has an exponential decay length that is long compared to other scale lengths of interest, such as the dephasing length or pulse decay due to energy depletion in laser wakefield acceleration.

Here, quasi-bound or leaky modes were found for azimuthally averaged density profiles by solving the Helmholtz equation for the field $E(r,z) = \mathcal{E}(r)e^{i\beta z}$ in cylindrical coordinates,

$$d^2\mathcal{E}/ds^2 + s^{-1}d\mathcal{E}/ds + (n^2 - \beta^2/k_0^2)\mathcal{E} = 0 \;, \tag{S.1}$$

where $k_0$ is the vacuum wave number, $\beta$ is the waveguide propagation wavenumber, $s = k_0 r$, and $n(r)$ is the refractive index profile derived from the azimuthally averaged plasma profile. To map out the quasi-bound modes for a given index profile $n(r)$, we repeatedly solve Eq. (S.1), scanning $\beta' = \beta/k_0$ and plotting $\eta(\beta') = (|\mathcal{E}_{vacuum}|^2 A)^{-1} \int_A |\mathcal{E}|^2 dA$, where the integral is over the waveguide cross section. The main peaks in $\eta(\beta')$ identify the quasi-bound modes, with the full width at half maximum $\Delta\beta$ of these peaks giving the axial ($z$) $1/e$ attenuation length of intensity $L_{1/e} = \Delta\beta^{-1}$.

Guided modes for the full 2D plasma profiles were calculated using a Beam Propagation Method (BPM) simulation [4]. A test mode overfilling the plasma density profile was input into the waveguide and propagated using a split-step Fourier method over more than a 100 Rayleigh lengths until all the high-order modes have decayed away. The fundamental guided mode is then propagated in the same manner and the peak intensity is sampled and fit to an exponential function to determine $L_{1/e}$.

## 3. Simulation of ultrashort Bessel beam pulse propagation and ionization

The simulations of higher order Bessel beam propagation through a pre-existing plasma column in Fig. 3(b) were performed using a 3D UPPE [5] implementation called YAPPE ('Yet Another Pulse Propagation Effort'). UPPE ('Unidirectional Pulse Propagation Equation') is a system of ordinary differential equations (ODEs) of the form

$$\frac{\partial}{\partial z} A_{k_x,k_y}(\omega,z) = iQ_{k_x,k_y}(\omega) 2\pi P_{k_x,k_y}(\omega,z) e^{-i\left(k_z - \frac{\omega}{v_g}\right)z} \;. \tag{S.2}$$

In Eq. (S.2), $A = A_{k_x,k_y}(\omega,z)$ is an auxiliary field related to the Fourier transform of the optical field by $E = Ae^{ik_z z}$. The spectrum of transverse spatial frequencies $(k_x, k_y)$ indexes a system of ordinary differential equations, which is solved using a GPU implementation of MATLAB's ODE45 function. $P_{k_x,k_y}(\omega,z)$ is the nonlinear polarization of the medium including (for our simulations in this paper): Kerr self-focusing, ionization losses, and a non-dispersive plasma response. The ionization rate is computed using an ADK module with neutral density saturation [6,7]. The perturbed index of refraction due to the $H_2$ neutral density gradient is also included as a part of the nonlinear polarization. The nonlinear refractive index ($n_2$) of hydrogen[8] is scaled with the spatiotemporally variable $H_2$ neutral density such that it goes to zero when all neutrals are ionized.

The higher order Bessel ($J_q$) beams were initialized as Laguerre-Gaussian $LG_{0q}$ modes with an applied conical (axicon) phase. The axicon phase shift was defined in YAPPE as $\Delta\varphi = -k_0 r \tan\gamma$ for central wavenumber $k_0$, distance from central axis $r$, and axicon base angle $\alpha = \gamma/2$. The beam was then propagated to its focus, where the simulation results in Fig. 4(b) of the main paper are reported. For that simulation, we launched a 2 mJ, 45 fs FWHM $LG_{0,16}$ pulse with waist $w_0 = 0.22$ mm, with an applied conical phase shift $\Delta\varphi$ for $\gamma = 3°$.

## References

[1] M. Takeda, H. Ina, and S. Kobayashi, "Fourier-transform method of fringe-pattern analysis for computer-based topography and interferometry," *J. Opt. Soc. Am.* **72**, 156 (1982).

[2] V. Dribinski, A. Ossadtchi, V. A. Mandelshtam, and H. Reisler, "Reconstruction of Abel-transformable images: The Gaussian basis-set expansion Abel transform method," *Rev. Sci. Instrum.* **73**, 2634 (2002).

[3] T. R. Clark and H. M. Milchberg, "Optical mode structure of the plasma waveguide," *Phys. Rev. E* **61**, 1954 (2000).

[4] K. Okamoto, *Fundamentals of optical waveguides* (Academic Press, 2005).

[5] M. Kolesik and J. V. Moloney, "Nonlinear optical pulse propagation simulation: From Maxwell's to unidirectional equations," *Phys. Rev. E* **70**, 11 (2004).

[6] V. S. Popov, "Tunnel and multiphoton ionization of atoms and ions in a strong laser field (Keldysh theory)," *Physics-Uspekhi* **47**, 855 (2004).

[7] S. V. Popruzhenko, V. D. Mur, V. S. Popov, and D. Bauer, "Strong Field Ionization Rate for Arbitrary Laser Frequencies," *Phys. Rev. Lett.* **101**, 193003 (2008).

[8] J. K. Wahlstrand, S. Zahedpour, Y.-H. Cheng, J. P. Palastro, and H. M. Milchberg, "Absolute measurement of the ultrafast nonlinear electronic and rovibrational response in $H_2$ and $D_2$," *Phys. Rev. A* **92**, 063828 (2015).